\documentclass[aps,prl,superscriptaddress,twocolumn,showpacs]{revtex4}
\usepackage{amssymb}
\usepackage{graphicx}
\usepackage{dcolumn}
\usepackage{longtable}
\newcommand{\lbar}{\ensuremath{\bar{\Lambda}}}
\newcommand{\pbar}{\ensuremath{\bar{p}}}
\newcommand{\gevc}{{\rm GeV}/c}

\newcommand{\degr}{^{\circ}}
\newcommand{\pip}{\ensuremath{\pi^+}}
\newcommand{\Mt}{m_\perp}
\newcommand{\Pt}{p_\perp}
\newcommand{\invyield}[0]{\ensuremath{\frac{1}{2\pi\Mt}\frac{d^2N}{d\Mt dy}}}
\newcommand{\invyieldof}[1]{\ensuremath
	{\frac{1}{2\pi\Mt}\frac{d^2N_{#1}}{d\Mt dy}}}
\newcommand{\dndy}[0]{\ensuremath{\frac{dN}{dy}}}
\newcommand{\dndyof}[1]{\ensuremath{\frac{dN_{#1}}{dy}}}
\newcommand{\asym}[3]{\ensuremath{#1^{+ #2}_{- #3}}}
\newcommand{\asymerr}[2]{\ensuremath{\mbox{}^{+ #1}_{- #2}}}
\newcommand{\sci}[2]{\ensuremath{#1 {\rm x} 10^{#2}}}

\newcommand{\Argonne}{Argonne National Laboratory, Argonne, IL 60439}
\newcommand{\BNL}{Brookhaven National Laboratory, Upton, NY 11973}
\newcommand{\UCR}{University of California Riverside, Riverside, CA
	92521}
\newcommand{\Columbia}{Columbia University, Nevis Laboratory, Irvington, NY
	10533}
\newcommand{\MIT}{Massachusetts Institute of Technology, Cambridge, MA
	02139}
\newcommand{\UIC}{University of Illinois at Chicago, Chicago, IL
	60607}
\newcommand{\Maryland}{University of Maryland, College Park, MD 20742}
\newcommand{\Yonsei}{Yonsei University, Seoul 120-749, South Korea}
\newcommand{\UoR}{University of Rochester, Rochester, NY 14627}

\begin{document}

\title{Antilambda production in Au+Au collisions at 11.7 $A$$\cdot$GeV/$c$}

\author{B.B.~Back}
\affiliation{\Argonne}
\author{R.R.~Betts}
\affiliation{\Argonne}
\affiliation{\UIC}
\author{J.~Chang}
\affiliation{\UCR}
\author{W.C.~Chang}
\affiliation{\UCR}
\author{C.Y.~Chi}
\affiliation{\Columbia}
\author{Y.Y.~Chu}
\author{J.B.~Cumming}
\affiliation{\BNL}
\author{J.C.~Dunlop}
\affiliation{\MIT}
\author{W.~Eldredge}
\affiliation{\UCR}
\author{S.Y.~Fung}
\affiliation{\UCR}
\author{R.~Ganz}
\affiliation{\UIC}
\author{E.~Garcia}
\affiliation{\Maryland}
\author{A.~Gillitzer}
\affiliation{\Argonne}
\author{G.~Heintzelman}
\affiliation{\MIT}
\author{W.F.~Henning}
\affiliation{\Argonne}
\author{D.J.~Hofman}
\affiliation{\UIC}
\author{B.~Holzman}
\affiliation{\UIC}
\author{J.H.~Kang}
\author{E.J.~Kim}
\author{S.Y.~Kim}
\author{Y.~Kwon}
\affiliation{\Yonsei}
\author{D.~McLeod}
\affiliation{\UIC}
\author{A.C.~Mignerey}
\affiliation{\Maryland}
\author{M.~Moulson}
\affiliation{\Columbia}
\author{V.~Nanal}
\affiliation{\Argonne}
\author{C.~Ogilvie}
\affiliation{\MIT}
\author{R.~Pak}
\affiliation{\UoR}
\author{A.Ruangma}
\affiliation{\Maryland}
\author{D.~Russ}
\affiliation{\Maryland}
\author{R.~Seto}
\affiliation{\UCR}
\author{P.J.~Stanskas}
\affiliation{\Maryland}
\author{G.S.F.~Stephans}
\affiliation{\MIT}
\author{H.~Wang}
\affiliation{\UCR}
\author{F.L.H.~Wolfs}
\affiliation{\UoR}
\author{A.H.~Wuosmaa}
\affiliation{\Argonne}
\author{H.~Xiang}
\affiliation{\UCR}
\author{G.H.~Xu}
\affiliation{\UCR}
\author{H.~Yao}
\affiliation{\MIT}
\author{C.M.~Zou}
\affiliation{\UCR}
\collaboration{The E917 Collaboration}
\noaffiliation

\date{\today}

\begin{abstract}
We present results from Experiment E917 for antilambda and
antiproton production in Au+Au collisions at 11.7 $A$$\cdot\gevc$. We
have measured invariant spectra and yields for both species in
central and peripheral collisions. We find that the \lbar/\pbar\ ratio
near mid-rapidity
increases from $\asym{0.26}{0.19}{0.15}$ in peripheral collisions to
$\asym{3.6}{4.7}{1.8}$ in central collisions, a value that is substantially
larger than current theoretical estimates.
\end{abstract}


\maketitle

Enhanced antimatter and strangeness production have both been proposed
as signatures of the phase transition from normal hadronic matter
to the quark-gluon plasma (QGP) \cite{RafMu82,Raf82,Koch86}.
The yields of strange antibaryons combine both of these signatures in
a single production channel, and may prove to be a
particularly sensitive probe for the formation of the QGP.  Of
particular interest is the ratio \lbar/\pbar\ \cite{RafMu82}, which reflects
the relative abundance of $\bar{s}$ quarks to light antiquarks.

However, both \lbar\ and \pbar\ are also produced when the
collision does  not form a QGP. Hadronic matter with sufficient 
rescattering can reach a chemical equilibrium containing the full
spectrum of hadrons.
Equilibrium chemical models of heavy-ion collisions are
generally able to reproduce the measured ratios of
total yields of particles such as pions and kaons.  
Such models also predict values for the ratio \lbar/\pbar\ in the 
range 0.15--0.9 at AGS beam energies \cite{RafDanos94,Bec00}.
A thermal model analysis that accounts for the measured
K$^+/\pi^+$ and K$^-$/K$^+$ ratios \cite{Welke98} gives 
a most probable value of \lbar/\pbar\ = 1, with 
unrealistic values yielding an upper limit of 2 for Au+Au 
collisions.

Hadronic cascade models which explicitly follow the
reactions and trajectories of hadrons in the collision zone
\cite{Stoecker97,Bass99} predict values for 
the \lbar/\pbar\ ratio of less than 1.
The absorption cross-section for \lbar-N processes is not well known 
at these energies, opening the
possibility for differential absorption of the two species to play a
role in determining the 
value of the ratio $\lbar/\pbar$ and its rapidity dependence. 
However, a study has shown that a reasonable interpretation
of this value cannot bring the predictions of cascade 
models above a ratio of $\sim 1$ \cite{Welke98}.
Recent data from E877 on the anti-flow of \pbar ~\cite{e877pbar}
are successfully described by the hadronic cascades 
that use the normal antiproton annihilation cross-section.

Enhanced antibaryon production may also be due to
many-body collisions between pions, 
{\it e.g.} $n\pi \rightarrow \pbar+p$ \cite{Rap00,Greiner00}.
The rate of antibaryon production can be estimated
from the inverse reaction $\pbar+p\rightarrow n\pi$.
Calculations by Rapp \cite{Rap00} and Greiner \cite{Greiner00}
indicate that a significant number of both $\pbar$ and $\lbar$ may come
from these processes, 
but it is not yet clear how to incorporate these many-body collisions into
current hadronic cascade calculations.

In summary, hadronic models with reasonable
inputs generally predict $\lbar/\pbar \leq 1 $. 
Thus the experimental detection of a \lbar/\pbar\ ratio 
significantly larger than 1 may signal the breakdown 
of these hadronic models, and possibly the onset of other
degrees of freedom in the system. 
Of particular interest is the centrality dependence of the
\lbar/\pbar\ ratio, which may discriminate \cite{Bleicher99}
between the various models attempting to describe antibaryon
production: QGP, hadronic scattering, and thermal models.
Canonical thermal models should be used in this comparison,
since the volume of the emitting system has been shown 
to affect the ratio of strange to non-strange
particles by up to a factor of four \cite{Hamieh00}.

There are indications of an abnormally
large value of the \lbar/\pbar\ ratio from other experiments.  
Experiment E859 reported a directly
measured  \lbar/\pbar\ ratio
of $2.9\pm0.9\mathrm{(stat)}\pm0.5\mathrm{(sys)}$ near mid-rapidity 
in central Si+Au collisions, at a beam energy 
of 14.6 $A$$\cdot\gevc$ \cite{GSFS}.
An indirect estimate of $\lbar/\pbar > 2.3$ (98\% C.L.) at y=1.6
for $\Pt \sim 0$ in central Au+Pb collisions was obtained 
at a beam energy of 11.5 $A$$\cdot\gevc$ by comparing 
inclusive antiproton yields from two different experients
(E864 and E878) with largely different acceptances \cite{E864paper}.

In this letter, we present a direct measurement of \lbar\ 
in Au+Au collisions near 
mid-rapidity and over an extended range in $\Pt$. 
We also measure the total \pbar\ spectrum, which includes contributions
from both direct \pbar\ production and \lbar\ decay. 
Using the \lbar\ measurement, we can estimate the contribution
to \pbar\ from \lbar\ decay and thus extract the yield of 
directly produced antiprotons ($\pbar_{direct}$). We find the ratio 
$\lbar/\pbar_{direct}$ to be large in central collisions, and
much lower in more peripheral collisions.  

The experiment was carried out using the magnetic tracking
spectrometer previously employed in AGS experiments E802,
E859, and E866 as described in detail in
Refs. \cite{nim,Chen98,kpkm,Back99}. An efficient second-level 
trigger was used in order to study the production of rare 
particles such as antiprotons. 
A $^{197}$Au beam with momentum 11.7 $A$$\cdot\gevc$ was incident on a 
1-g/cm$^2$ thick Au target.  The movable spectrometer 
subtended polar angles from $19\degr$ to
$34\degr$.  Magnetic field settings of $\pm 0.4$ T were used, for
which the momentum resolution of the spectrometer, $\delta p/p$, ranges from
1\% to 2\%. At these angle and field settings,
we have acceptance for \pbar\ and \lbar\ in the rapidity range $1.0 <
y < 1.4$, where mid-rapidity is $y=1.6$ at this beam energy. 
Particle identification is based on the measured momentum combined
with a time-of-flight (TOF) wall with a resolution of
135~ps situated 6~m from the target.  

We define centrality classes based on the fraction of the
total interaction cross-section ($\sigma_{tot} = 6.8~{\rm barns}$ 
\cite{Geer95}). Using the total kinetic energy of spectator particles
measured in a zero-degree  calorimeter, the data were divided into two
centrality classes, called ``Central'' and ``Peripheral'', containing
$0\%-12\%$ and $12\%-77\%$ of $\sigma_{tot}$, respectively. We also
report a ``Minimum Bias'' data set without centrality cuts.

The reconstruction, particle identification and tracking algorithms have been
described in detail elsewhere \cite{kpkm}.  
Antiprotons are accepted up to momentum 
$2.9~\gevc$, above which the $3\sigma$ identification bands of the K$^-$
and $\pbar$ time of flight spectrum begin to overlap.  
A lower momentum cutoff of $0.5~\gevc$ was necessitated by 
hadronic interaction rates in the spectrometer material. 
Pions used to reconstruct \lbar's were required to have momenta 
between 0.35--1.8 $\gevc$.

A firm understanding of the background in the TOF-identified $\pbar$ sample
is required for the measurement of the $\pbar$ spectrum. 
Spectra of the time-of-flight residuals 
$\Delta t = (t_{meas} - t_{expected})/\sigma$ were created 
in order to estimate these backgrounds in the $\pbar$ sample. 
Here $t_{meas}$ and $t_{expected}$ are the
measured and expected flight times of $\pbar$'s to the TOF wall, and
$\sigma$ is a particle-by-particle estimate of the timing resolution
based on the TOF wall resolution combined with a Monte Carlo
simulation of the tracking in the spectrometer.
The $\Delta t$ distributions are parameterized by a Gaussian
for the true $\pbar$ signal and a background with two components: an
exponentially falling contribution from the tails of K$^-$ and $\pi^-$
bands, and a flat random contribution due to misidentified 
particles. The backgrounds
extracted range from 15\% to 75\% of the total signal, with 
the largest backgrounds at the extremes of the momentum ranges.
Rapidity-transverse mass bins for which the background
fraction exceeds 50\% have been excluded from further analysis.

\begin{figure}
\includegraphics[width=3in]{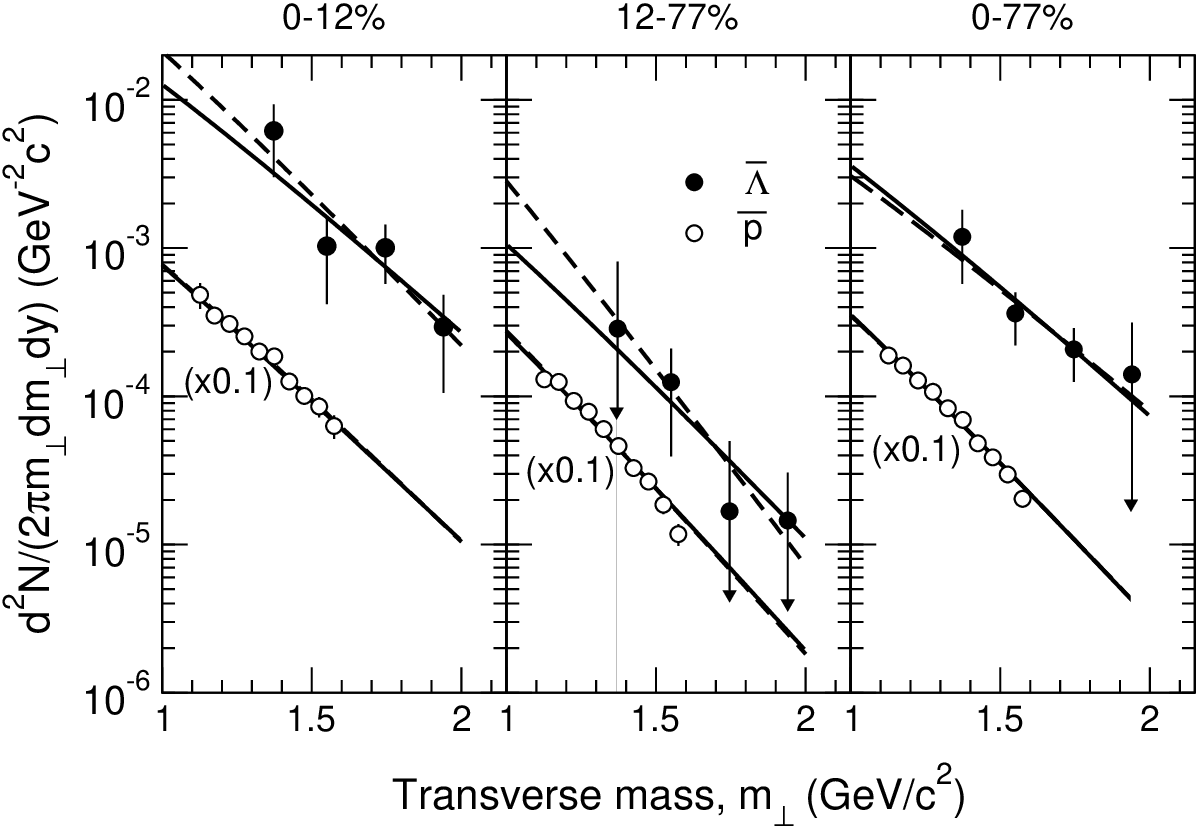}
\caption{Invariant spectra as a function of transverse mass
for \lbar\ (solid circles) and \pbar\ (open circles, multiplied by 0.1 for 
clarity) in the three centrality classes. 
The error bars include a 2\% point-to-point systematic error due to
acceptance corrections.
The solid lines are from the simultaneous fit as described in the text and
the dashed lines are from independent fits to \lbar\ and \pbar\ 
spectra shown for comparison. Solid and dashed fit lines overlap for the
\pbar\ spectra.}
\label{fig:spectra}
\end{figure}

The counting statistics in determining the background are included 
in the reported statistical error bars. We
estimate an additional systematic error in the extracted \pbar\ yield
due to background corrections and acceptance uncertainties to be 10\%,
and the error in the \pbar\ inverse slope, arising mostly from acceptance
uncertainties, to be 5\%. 
The measured \pbar\ invariant spectra are shown as open circles 
in Fig.~\ref{fig:spectra} as 
a function of transverse mass ($\Mt = \sqrt{\Pt^2 + m_0^2}$), 
along with independent fits (dashed lines) of a
Boltzmann form, $B(T)$, {\it i.e.}
\begin{eqnarray*}
\invyield & = & \dndy B(T) \nonumber \\
          & = &
	\dndy \frac{1}{2 \pi } \Mt 
	\frac{\exp(-(\Mt-m_0)/T)}{m_0^2T+2m_0T^2+2T^3} . 
\label{eqn:boltz} 
\end{eqnarray*}
The parameters of the fit to the total \pbar\ spectra
($dN_{\pbar_{total}}/dy$, $T_{\pbar_{total}}$) are given in columns two
and three of Table \ref{tab:double_fit_params}.
These spectra include contributions from both 
direct production ($\pbar_{direct}$) and anti-hyperon decay
($\pbar_{decay}$), which we now consider. 

\begin{figure}
\includegraphics[width=3in]{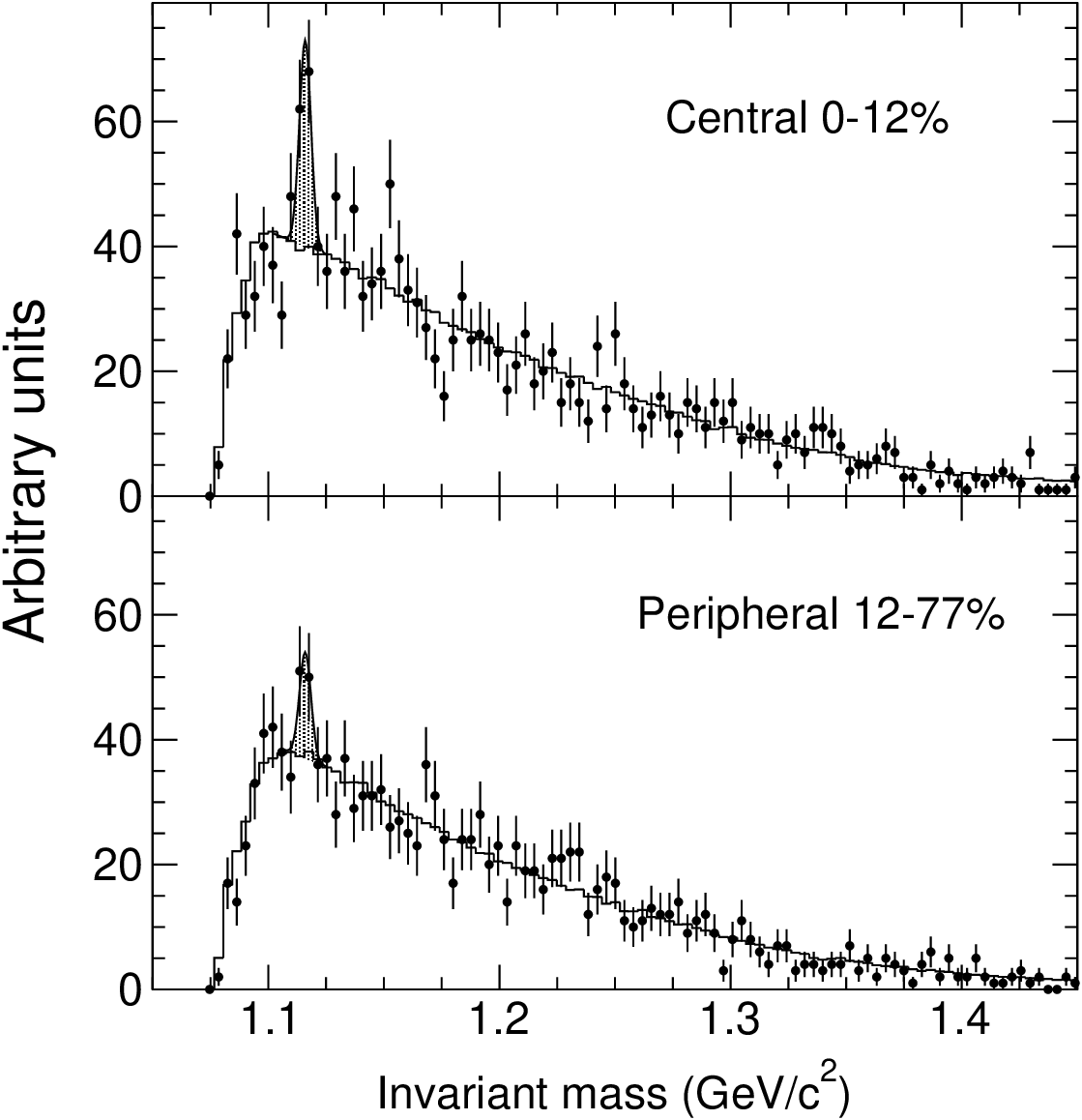}
\caption{The \lbar\ invariant mass spectra for central and peripheral 
collisions for the data in Fig.~{\protect \ref{fig:spectra}}. 
Solid points are measured data, and 
the histogram is the normalized mixed-event background after all 
corrections have been applied. The \lbar\ mass peak is shaded.}
\label{fig:lbar_minv}
\end{figure}

We reconstruct \lbar's through direct measurement of 
$(\pbar,\pip)$ pairs. This decay channel has a branching
ratio of 63.9\% \cite{PDB}.  The spectrum of reconstructed 
invariant mass for accepted $(\pbar,\pip)$ pairs is shown 
in Fig.~\ref{fig:lbar_minv}.
The experimental resolution for reconstructing \lbar's is found by
initially reconstructing the $\Lambda$ invariant mass spectrum in the
same data set. The peak of the $\Lambda$ spectrum is found to be at 
$1.1161\pm0.0004~{\rm GeV}/c^2$,
consistent with the accepted value of $m_\Lambda=1.1157~{\rm GeV}/c^2$. 
The Gaussian width of the peak is $\sigma=1.3~{\rm MeV}/c^2$, which is
attributable to experimental resolution.  To define \lbar\ candidates,
a cut of $\pm 3\sigma$ is taken around the nominal $m_\Lambda$.
To estimate the background under the \lbar\ peak, we have constructed
a mixed-event background, shown as the solid histogram in 
Fig.~\ref{fig:lbar_minv}.  Pairs are constructed by selecting random
\pbar's and \pip's from the set of all pairs in the centrality
class being considered.  The background is normalized to the region
outside the \lbar\ peak, defined as more than $6\sigma$ from 
$m_\Lambda$.  Both real and mixed pairs were included only if 
their opening angle was greater than 15~mrad. This removes most of the 
inefficiency for detecting close tracks. The remaining two-track
inefficiency is estimated by comparison with the mixed-event sample. 
The correction is then applied on a pair-by-pair basis in 
generating the background invariant mass spectrum as well
as to the signal's final $\Mt$ spectra. This correction has 
less than a 5\% effect on the yields. 

We further correct the estimated background for the residual 
correlations arising from the presence of particles from signal pairs
in the mixed-event background.  Pairs in the mixed-event background 
with a member that came from the signal region are assigned a 
weight equal to the background fraction.
This must be done in an iterative fashion; four
iterations are sufficient to attain convergence.  This correction
increases the extracted \lbar\ yields by approximately 15\%. From
the small remaining discrepancy between the shape of the mixed-event
sample and the actual sample, we estimate the systematic error on the
measured \lbar\ yield arising from background subtraction to be
15\%.  The \lbar\ invariant spectra are shown as solid circles 
in Fig.~\ref{fig:spectra} as
a function of transverse mass in all centrality classes. 

Since the \pbar\ spectra have a strong contribution from \lbar\ decays,
it is advantageous to perform a simultaneous fit to both spectra, in
order to obtain the most accurate measure of the \lbar/\pbar\ ratio.
We have used a  Monte Carlo study to parameterize the
distribution of \pbar's originating from \lbar\ decay, as a function
of the characteristics of the original \lbar\ spectrum. For a \lbar\
spectrum of Boltzmann form with inverse slopes $T$ in the range 
$150~{\rm MeV} < T < 300~{\rm MeV}$, we
find that the resulting spectrum of \pbar's within
the E917 spectrometer is also Boltzmann in shape, with parameters:
\begin{eqnarray*}
T_{\pbar_{decay}} & = & (0.834 \pm 0.006) T_{\lbar} + (0.004 \pm
0.001)~{\rm (GeV) } \\
\dndyof{\pbar_{decay}} & = & 0.639~\dndyof{\lbar} .
\end{eqnarray*}
The errors reflect the quality of the Boltzmann fits to
the $\pbar_{decay}$ spectra, and result in a negligible systematic
error to the final results.

The \lbar\ and \pbar\ transverse mass spectra were therefore fit
simultaneously using the functions
\begin{eqnarray*}
\invyieldof{\lbar} & = & \dndyof{\lbar} B(T_{\lbar}), 
\label{eqn:firstsimul} \mathrm{~and}
\\
\invyieldof{\pbar} & = & \dndyof{\lbar} \left(
	\frac{B(T_{\pbar_{direct}})}{\lbar/\pbar_{direct}}
	+  0.639 B(T_{\pbar_{decay}}) \right).
\label{eqn:secondsimul}
\end{eqnarray*}
From these fits, shown in Fig.~\ref{fig:spectra},
we obtain values of the four free parameters
$T_{\pbar_{direct}}$, $dN_{\lbar}/dy$, $T_{\lbar}$, and
$\lbar/\pbar_{direct}$, listed in Table \ref{tab:double_fit_params}.
The \lbar/\pbar\ ratio and \lbar\ yield obtained in
this manner are consistent with those obtained from separate fits to
the \lbar\ and \pbar\ spectra, but the precision of the simultaneous
fitting method is superior due to the additional constraint on the
\lbar\ inverse slope provided by the shape of the \pbar\ spectrum.

In summary, we have made a direct measurement of \lbar\ production in Au+Au
collisions at the AGS. We find that the \lbar\ production increases sharply
with centrality, from $\dndy = \sci{\asym{1.2}{0.7}{0.6}}{-3}$ in
peripheral collisions to $\dndy = \sci{\asym{19}{4}{5}}{-3}$ in
central collisions. The total \pbar\ production is of
comparable magnitude to the \lbar\ yield, with a substantial fraction
of the measured \pbar\ signal attributable to decay products from
the \lbar.

\begin{figure}
\includegraphics[width=3in]{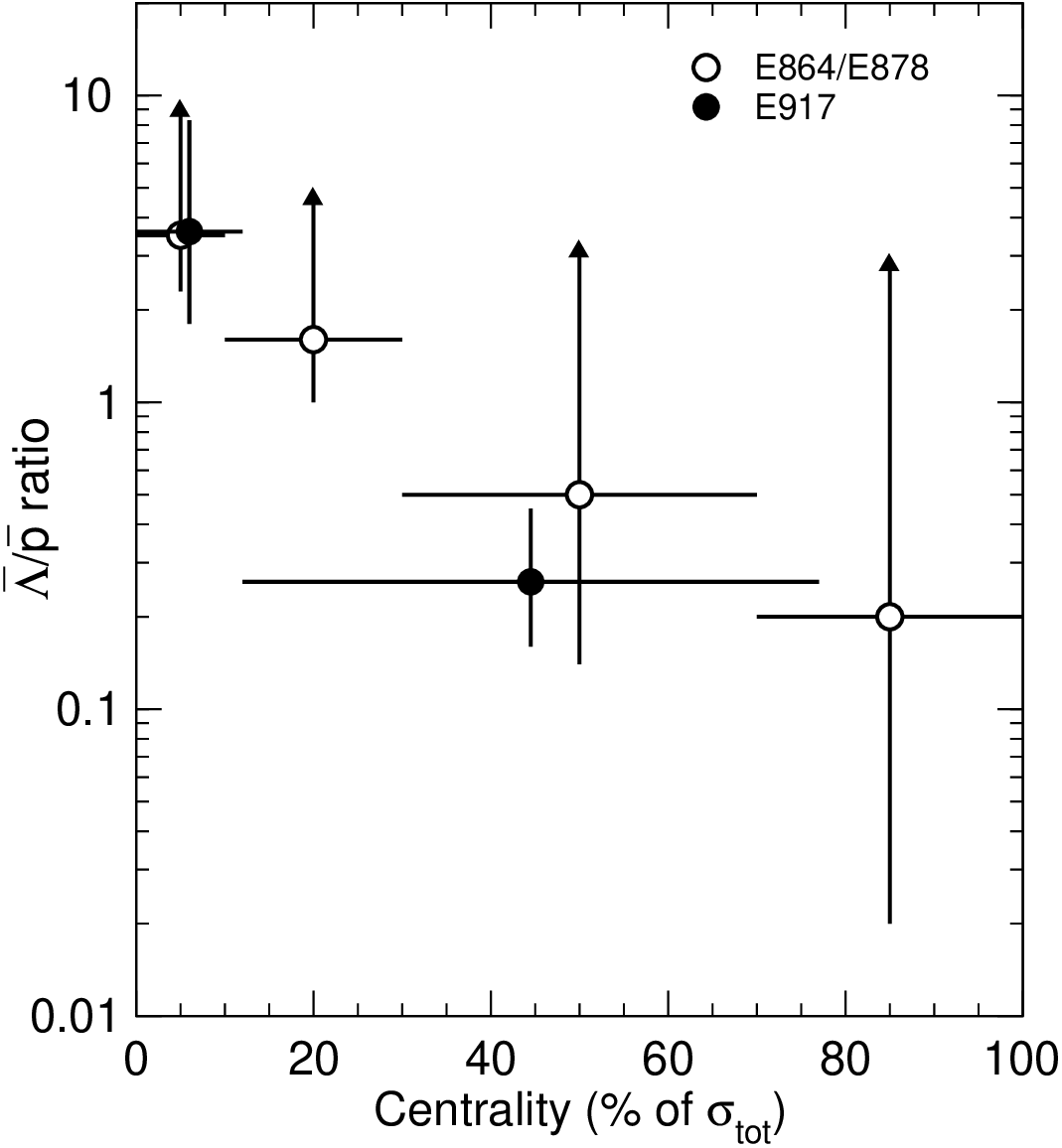}
\caption{The \lbar/\pbar\ ratio is shown as a function of the centrality
given in terms of the percentage of the total interaction cross section,
$\sigma_{tot}$. The E917 data are integrated over all 
$\Pt$ and $1.0<y<1.4$ and are shown as solid circles with 1$\sigma$
statistical error bars for the \lbar/\pbar\ ratio. The E864/E878 
indirect results {\protect \cite{E864paper}} are at $\Pt \sim 0$
and y=1.6 and are shown as most probable values 
with the 98\% confidence level of the minimum value indicated 
by the lower error bar. Upper limits are not given.
Horizontal bars indicate bin widths.}
\label{fig:impact}
\end{figure}

We find that the ratio \lbar/\pbar\ near mid-rapidity 
increases from 	$\asym{0.26}{0.19}{0.15}$ in peripheral collisions to
$\asym{3.6}{4.7}{1.8}$ in central collisions.
When combined with the prior observations of elevated ratios 
in central Si+Au collisions \cite{GSFS} and the indirect evidence
from experiments E864 and E878 \cite{E864paper}, the present
results indicate that \lbar\ production is larger than 
direct \pbar\ production in central collisions---and not in 
peripheral collisions---at AGS beam energies. 
This effect is clearly shown in Fig.~\ref{fig:impact}.
The mechanism responsible for the large \lbar/\pbar\ ratio is 
currently unknown.  

\begin{acknowledgments}
This work is supported by the U.S. Department of Energy under contracts 
with ANL (No.~W-31-109-ENG-38), BNL (No.~DE-AC02-98CH10886), MIT 
(No.~DE-AC02-76ER03069), UC Riverside (No.~DE-FG03-86ER40271), UIC 
(No.~DE-FG02-94ER40865) and the University of Maryland
(No.~DE-FG02-93ER40802), the National Science Foundation under 
contract with the University of Rochester (No.~PHY-9722606), and 
the Ministry of Education and KOSEF (No.~951-0202-032-2) in Korea.
\end{acknowledgments}

\widetext

\begin{table*}
\caption{Parameters of an independent fit to the
total measured \pbar\ spectrum
($dN_{\pbar_{total}}/dy$, $T_{\pbar_{total}}$) and
a simultaneous fit (the remaining values) to \pbar\ and \lbar\
spectra, for the three centrality bins. See text for an explanation of
the parameters. Errors are statistical followed by systematic.} 
\begin{tabular}{ccc|cccc}
\toprule
Bin &
$dN_{\pbar_{total}}/dy$ &
$T_{\pbar_{total}}$ &
$T_{\pbar_{direct}}$ &
$dN_{\lbar}/dy$ & 
$T_{\lbar}$ & 
$\lbar/\pbar_{direct}$ \\
    &
($\sci{}{-3}$) &   
(GeV) &
(GeV) &
($\sci{}{-3}$) &
(GeV) &    \\
\hline
0-12\%
        & $17.7 \pm 0.5 \pm 1.8$
        & $0.200 \pm 0.008 \pm 0.010$
	& 0.23 \asymerr{0.15}{0.06}\asymerr{0.01}{0.01} 
	& 19 \asymerr{4}{5}\asymerr{3}{2}
	& 0.22 \asymerr{0.04}{0.03}\asymerr{0.01}{0.01}
	& 3.6 \asymerr{4.7}{1.8} \asymerr{2.7}{1.1}\\
12-77\%
        & $5.5 \pm 0.1 \pm 0.6$
        & $0.175 \pm 0.004 \pm 0.008$
	& 0.18 \asymerr{0.02}{0.01}\asymerr{0.01}{0.01}
        & 1.2 \asymerr{0.7}{0.6}\asymerr{0.2}{0.2}
	& 0.19 \asymerr{0.06}{0.03}\asymerr{0.01}{0.01}
        & 0.26 \asymerr{0.19}{0.15} \asymerr{0.5}{0.4}\\
0-77\%
        & $7.4 \pm 0.1 \pm 0.7$
        & $0.185 \pm 0.004 \pm 0.009$
        & 0.183 \asymerr{0.017}{0.014}\asymerr{0.014}{0.003}
        & 5.3 \asymerr{1.1}{1.1}\asymerr{1.0}{1.0}
	& 0.218 \asymerr{0.026}{0.020}\asymerr{0.020}{0.004}
        & 1.3 \asymerr{0.6}{0.4}\asymerr{0.6}{0.3}\\
\botrule
\end{tabular}
\label{tab:double_fit_params}
\end{table*}


\begin{thebibliography}{99}
\bibitem{Stoecker97} Exact values for the \lbar/\pbar$_{direct}$
ratio at AGS energies predicted by transport models have not been published.
Review articles on transport models {\protect \cite{Bass99}} state that
the models underpredict by a factor of three to five the measured
Si + Au ratio of \lbar/\pbar$ =2.9 \pm 0.9 \pm 0.5$.
\end{thebibliography}
\end{document}